\documentclass[twocolumn]{aastex62} 
\usepackage{subfigure}
\usepackage{url}
\usepackage{graphicx,color}
\usepackage{xcolor}
\usepackage{gensymb}

\usepackage{hyperref}

\newcommand{\FeIX}{\ion{Fe}{9}}
\newcommand{\FeXII}{\ion{Fe}{12}}
\newcommand{\FeXIV}{\ion{Fe}{14}}
\newcommand{\FeXVI}{\ion{Fe}{16}}
\newcommand{\FeXVIII}{\ion{Fe}{18}}
\newcommand{\FeXXI}{\ion{Fe}{21}}
\newcommand{\HeII}{\ion{He}{2}}

\DeclareMathAlphabet{\mathitbf}{OML}{cmm}{b}{it}
\DeclareMathAlphabet{\mathf}{OML}{cmm}{c}{sl}

\newcommand{\kms}{km~s$^{-1}$}

\shorttitle{Hi-C Flare 129 \AA\ observations of fine-scale coronal energy release}
\shortauthors{Tiwari et al.}

\begin{document}

\title{A sudden fine-scale bright kernel captured by {\it Hi-C Flare} in 11 MK emission during an M1.6-class solar flare's post-maximum phase}

\correspondingauthor{Sanjiv K. Tiwari}
\email{tiwari@lmsal.com}

\author[0000-0001-7817-2978]{Sanjiv K. Tiwari}
\affiliation{Lockheed Martin Solar and Astrophysics Laboratory, 3251 Hanover Street, Bldg. 203, Palo Alto, CA 94306, USA}
\affiliation{Bay Area Environmental Research Institute, NASA Research Park, Moffett Field, CA 94035, USA}

\author[0000-0001-7620-362X]{Navdeep K. Panesar}
\affiliation{Lockheed Martin Solar and Astrophysics Laboratory, 3251 Hanover Street, Bldg. 203, Palo Alto, CA 94306, USA}
\affiliation{SETI Institute, 339 Bernardo Ave, Mountain View, CA 94043, USA}

\author[0000-0002-5691-6152]{Ronald L. Moore}
\affiliation{NASA Marshall Space Flight Center, Mail Code ST 13, Huntsville, AL 35812, USA}
\affil{Center for Space and Aeronomic Research, The University of Alabama in Huntsville, Huntsville, AL 35805, USA}

\author[0000-0002-6172-0517]{Sabrina L. Savage} 
\affiliation{NASA Marshall Space Flight Center, Mail Code ST 13, Huntsville, AL 35812, USA}

\author[0000-0002-5608-531X]{Amy R. Winebarger} 
\affiliation{NASA Marshall Space Flight Center, Mail Code ST 13, Huntsville, AL 35812, USA}

\author[0000-0002-7219-1526]{Genevieve D. Vigil} 
\affiliation{NASA Marshall Space Flight Center, Mail Code ST 13, Huntsville, AL 35812, USA}

\author[0000-0002-9690-8456]{Juraj Lörinčík}
\affiliation{Bay Area Environmental Research Institute, NASA Research Park, Moffett Field, CA 94035, USA}
\affiliation{Lockheed Martin Solar and Astrophysics Laboratory, 3251 Hanover Street, Bldg. 203, Palo Alto, CA 94306, USA}

\author[0000-0002-4980-7126]{Vanessa Polito}
\affiliation{Lockheed Martin Solar and Astrophysics Laboratory, 3251 Hanover Street, Bldg. 203, Palo Alto, CA 94306, USA}

\author[0000-0002-8370-952X]{Bart De Pontieu}
\affiliation{Lockheed Martin Solar and Astrophysics Laboratory, 3251 Hanover Street, Bldg. 203, Palo Alto, CA 94306, USA}
\affil{Rosseland Centre for Solar Physics, University of Oslo, P.O. Box 1029 Blindern, NO–0315 Oslo, Norway}
\affil{Institute of Theoretical Astrophysics, University of Oslo, P.O. Box 1029 Blindern, NO–0315 Oslo, Norway}

\author[0000-0001-9638-3082]{Leon Golub} 
\affiliation{The Center for Astrophysics | Harvard and Smithsonian, 60 Garden St., Cambridge, MA 02138, USA}

\author[0000-0003-1057-7113]{Ken Kobayashi} 
\affiliation{NASA Marshall Space Flight Center, Mail Code ST 13, Huntsville, AL 35812, USA}

\author[0000-0002-7139-6191]{Patrick Champey}
\affiliation{NASA Marshall Space Flight Center, Mail Code ST 13, Huntsville, AL 35812, USA}

\author[0000-0002-4498-8706]{Jenna Samra}
\affiliation{The Center for Astrophysics | Harvard and Smithsonian, 60 Garden St., Cambridge, MA 02138, USA}

\author[0009-0002-8726-6196]{Anna Rankin} 
\affiliation{University of Lancashire, Preston, PR1 2HE, UK}

\author[0000-0002-1025-9863]{Robert W. Walsh} 
\affiliation{University of Lancashire, Preston, PR1 2HE, UK}

\author[0000-0001-5243-7659]{Crisel, Suarez}
\affiliation{The Center for Astrophysics | Harvard and Smithsonian, 60 Garden St., Cambridge, MA 02138, USA}
\affiliation{Vanderbilt University, 2201 West End Ave., Nashville, TN 37235, USA}

\author[0000-0002-4103-6101]{Christopher S. Moore}
\affiliation{The Center for Astrophysics | Harvard and Smithsonian, 60 Garden St., Cambridge, MA 02138, USA}

\author[0000-0002-4691-1729]{Adam R. Kobelski}
\affiliation{NASA Marshall Space Flight Center, Mail Code ST 13, Huntsville, AL 35812, USA}

\author[0000-0003-4739-1152]{Jeffrey, W. Reep}
\affiliation{Institute for Astronomy, University of Hawai'i at Mānoa, Pukalani, HI 96768, USA}

\author[0000-0002-1992-7469]{Charles, Kankelborg}
\affiliation{Montana State University, Culbertson Hall, 100, Bozeman, MT 59717, USA}

\graphicspath{{Figures/}}     

\begin{abstract}
On April 17, 2024, the third successful Hi-C sounding rocket flight, {\it Hi-C Flare}, recorded coronal images
in Fe XXI 129 \AA\ emission from 11 MK plasma during the post-maximum phase of an M1.6-class solar flare, achieving unprecedented spatial {($\sim$300 km)} and temporal (1.3 s) resolutions. 
The flare started at 21:55UT, peaked at 22:08UT, and lasted $\sim$40 minutes. Hi-C observed for over five
minutes (22:15:45 to 22:21:25), starting roughly eight minutes after flare maximum. 
A sudden compact bright burst---875$\pm$25 km wide, lasting 90$\pm$1.3 s, exhibiting a proper motion of  $\sim$50 \kms, and splitting into two toward the end---occurs near the foot of some post-flare loops. Its size and brightness are reminiscent of flare-ribbon kernels during a flare’s rapid rise phase, kernels marking sites of sudden heating and hot plasma upflow, making its occurrence during the late phase surprising. Such isolated brightenings in a flare's post-maximum phase are rare, and have not been previously reported. 
 The kernel was detected in all SDO/AIA channels. Its 1600 \AA\ light curve peaked $\sim$50 s earlier than its 131 \AA\ light curve, similar to that of flare-ribbon kernels, albeit with a smaller delay of $\sim$25 s, during the impulsive phase of the flare.
In SDO/HMI magnetograms, the kernel sits in unipolar positive magnetic flux near an embedded clump of negative flux. Although localized magnetic reconnection within the kernel {(a microflare)} cannot be ruled out for its cause, the observations favor the localized brightening being an isolated, exceptionally late flare-ribbon kernel, resulting from an exceptionally late burst of the flare's coronal reconnection. 
\end{abstract}


\section{Introduction} \label{sec:intr}

The third successful sounding-rocket flight of the High-Resolution Coronal Imager \citep[Hi-C 3, namely `Hi-C Flare':][]{vigi25} took coronal extreme ultra-violet (EUV) images of NOAA active region (AR) 13645 in 129 \AA\ (dominated by Fe XXI emission) with unprecedented spatial and temporal resolutions ({$\sim$300 km}, 1.3 s). The data was collected for over five minutes, during the period of 22:15:41 -- 22:21:25 UT on April 17, 2024, near solar disk center (AR position: S08E23). The Hi-C Flare captured a segment of the decay phase of a M1.6-class solar flare, about 8 minutes after the flare maximum. The images show a small-scale localized outstanding transient brightening at an outer edge of the flare. 
This type of isolated, small-scale, flare-kernel-like brightening during a flare's post-maximum phase has not been reported in previous coronal EUV observations.

Solar flares are one of the most complex challenges in astrophysics, with each new observation testing and advancing our understanding. Solar flares are thought to be powered by magnetic reconnection in stressed magnetic fields \cite[e.g.,][]{shib95,prie00,moor01,zwei09,josh11Bhuwan,benz17}. Many aspects of the reconnection process remain poorly understood — for example, the mechanisms of energy transport, the evolution of mass flows during the flare, and how a significant portion of the magnetic energy is converted into accelerated non-thermal particles.
Solar flares emit radiation across a broad range of the electromagnetic spectrum \citep{flet11}. By their peak 1--8 \AA\ soft X-ray flux observed by GOES, solar flares are categorized into different classes. 
The energy released in different flare classes can be considered proportional to their peak GOES 1--8 \AA\ flux. 
X-class flares are usually the most energetic, often releasing magnetic energy on the order of 10$^{32}$ erg \citep{ems05}, but can reach up to  10$^{33}$ erg \citep{ems12,aula13,tori17} or even higher \citep{saku22}.  M-, C-, B-, and A-class flares release approximately 10$^{31}$, 10$^{30}$, 10$^{29}$, and 10$^{28}$ erg, respectively \citep[e.g., ][Appendix C]{tiw09a}.

{The flare emission is predominantly from hot flare loops with temperatures exceeding 10 MK, which are anchored in bright chromospheric ribbons \citep{masu94,warr01,youn13,grah15}}. As per the standard flare model, also known as CSHKP model \citep{carm64,stur66,hira74,kopp76}, these ribbons mark the footpoints of the reconnected magnetic field lines and are sites of intense energy deposition via accelerated particles,  magnetic wave dissipation,  and thermal conduction from the corona.  The energy input leads to the evaporation of chromospheric plasma, which fills the newly reconnected coronal loops, forming the hot, dense structures observed in EUV and soft X-rays during the gradual phase of flares \citep{benz17}.

Because flare emission is primarily associated with hot flare loops that have temperatures exceeding 10 MK, Hi-C flare observations at 129 \AA, with sensitivity to emission of plasma at temperatures over 10 MK plasma, are well able to capture such events with unprecedented high spatial and temporal resolutions.
In the rapid rise phase of a solar flare, ribbon kernels can appear as localized brightenings that may exhibit apparent motions along ribbons driven by slipping magnetic reconnection \citep[e.g.,][]{janv13,sobo16,dudi16}. This initial brightening in flare loop feet provides crucial insights into the fundamental processes driving solar flares \citep{benz17}.

While the rapid rise, main, and decay phases of solar flares have been the focus of numerous observational studies \citep[][and references therein]{macc79,flet11,benz17,hino19}, the occurrence of flare-ribbon kernels during the post-maximum phase has rarely been reported \citep{czay99,sobo16}. One of the lesser-known topics in flare physics is the mechanism by which flares continue to be heated during their post-maximum phase \citep{czay99,czay01}. Proposed explanations for such heating include prolonged magnetic reconnection \citep{moor80,warr06_flare,reev07,kuha17}, compression from strong downflows \citep{reev17,unve23}, and shock heating \citep{yoko01}, or some combination of these. 
In this work, we report the observation of a prospective isolated flare-ribbon kernel during the post-maximum phase of a solar flare.

\begin{figure*}
	\centering
	\includegraphics[trim=7.8cm 0cm 8cm 0.3cm,clip,width=0.85\textwidth]{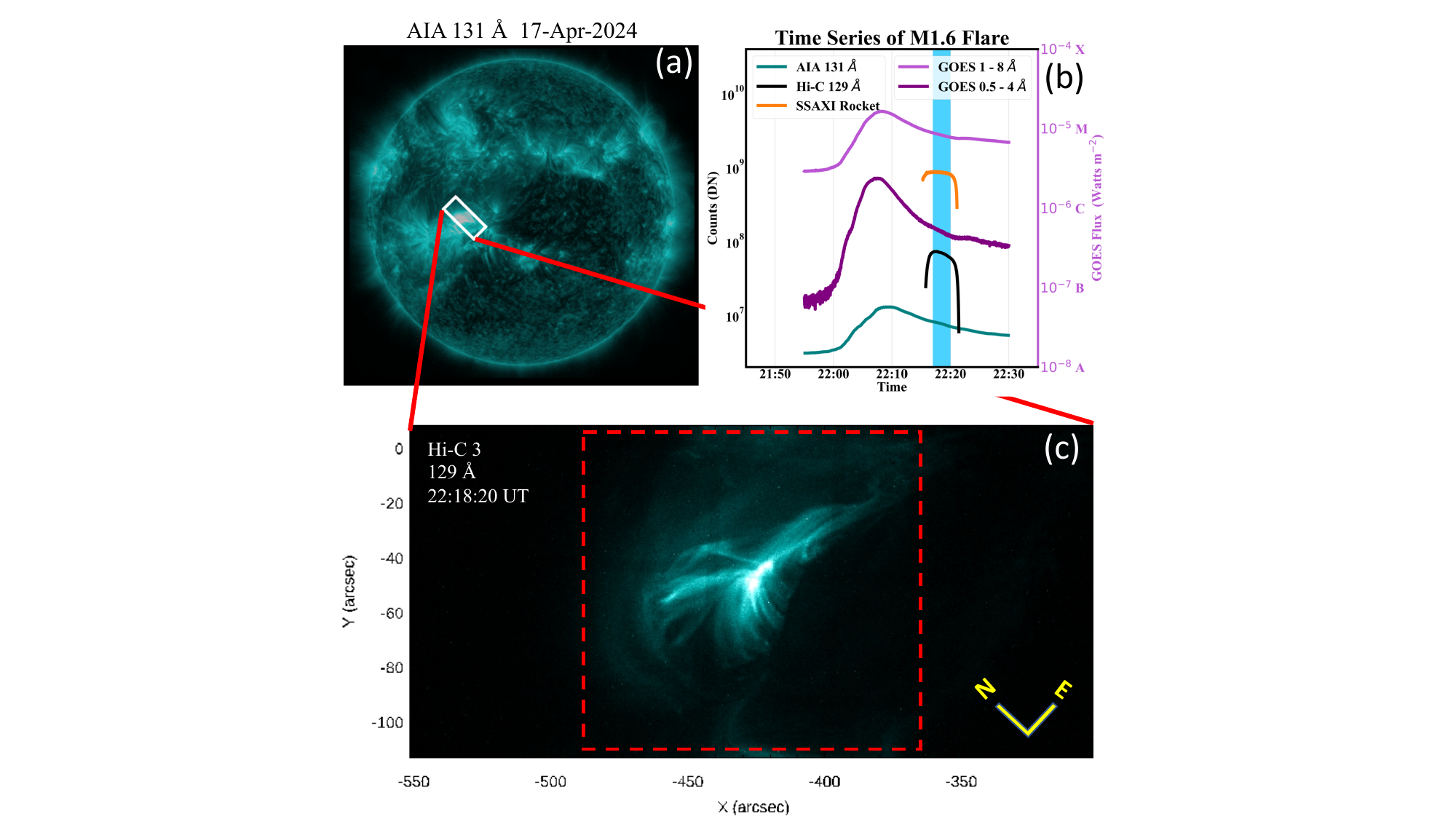}
	\caption{Context image of Hi-C Flare observations of NOAA AR 13645 at the solar disk position S08E23. Panel (a) shows full disk solar image from SDO/AIA in 131 \AA\ channel. A white box roughly outlines the region observed by Hi-C for over five minutes, a frame of which is shown in panel (c). Panel (b) shows GOES, AIA, Hi-C, and SSAXI-Rocket light curves of the observed flare. The shade in cyan color outlines the timing of the main phase of Hi-C observations. Panel (c) shows an image frame with the full field of view (FOV) observed by Hi-C Flare in 129 \AA. The dashed red box outlines the region with most activities and used for further analysis. Faint grid structure on Hi-C image is artefact of focal plane filter shadow that still remains in the data. The directions of solar north and east are indicated in the lower-right corner of panel (c).
	}
	\label{fig1}
\end{figure*}

\section{Data and Methods} \label{data}
The five minutes of Hi-C Flare observations, obtained at a cadence of 1.3 s and a spatial resolution of {$\sim$300} km, were complemented by the Helioseismic Magnetic Imager \citep[HMI:][]{scho12} and Atmospheric Imaging Assembly \citep[AIA:][]{leme12} onboard SDO \citep{pesn12}. 

{A degradation in the instrument’s spatial resolution from previous flights is expected as a result of the mirror re-coating process. The resolution limit of the instrument is currently estimated to be $\sim$0.45” and more detailed information on the actual spatial resolution of Hi-C 3 will be provided in the forthcoming instrument paper \citep{vigi25}. }

\begin{figure*}\centering
	\includegraphics[trim=0.5cm 0.8cm 0.55cm 1.2cm,clip,width=0.85\textwidth]{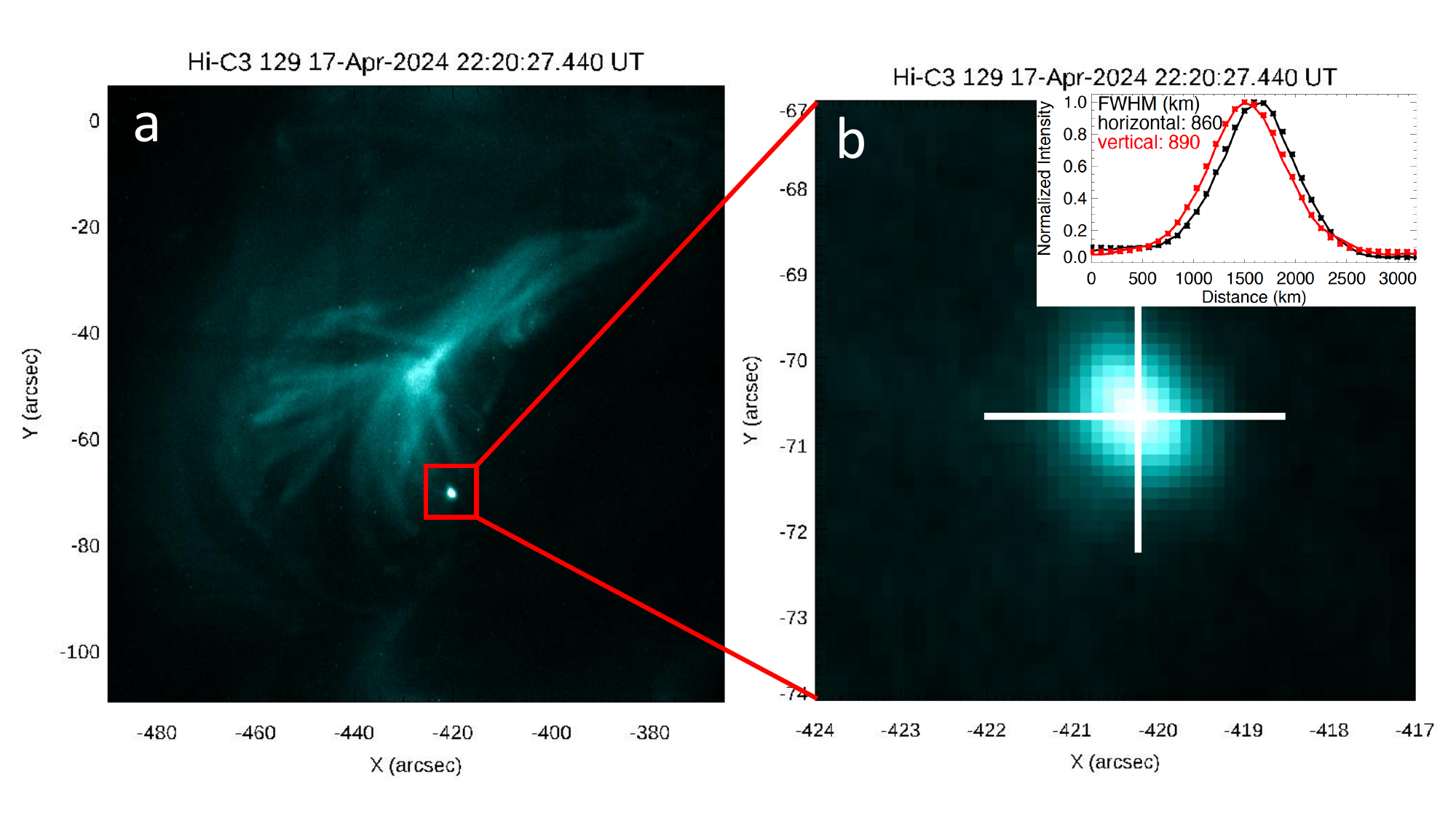}
\includegraphics[trim=1.7cm 0cm 7.1cm 0cm,clip,width=0.85\textwidth]{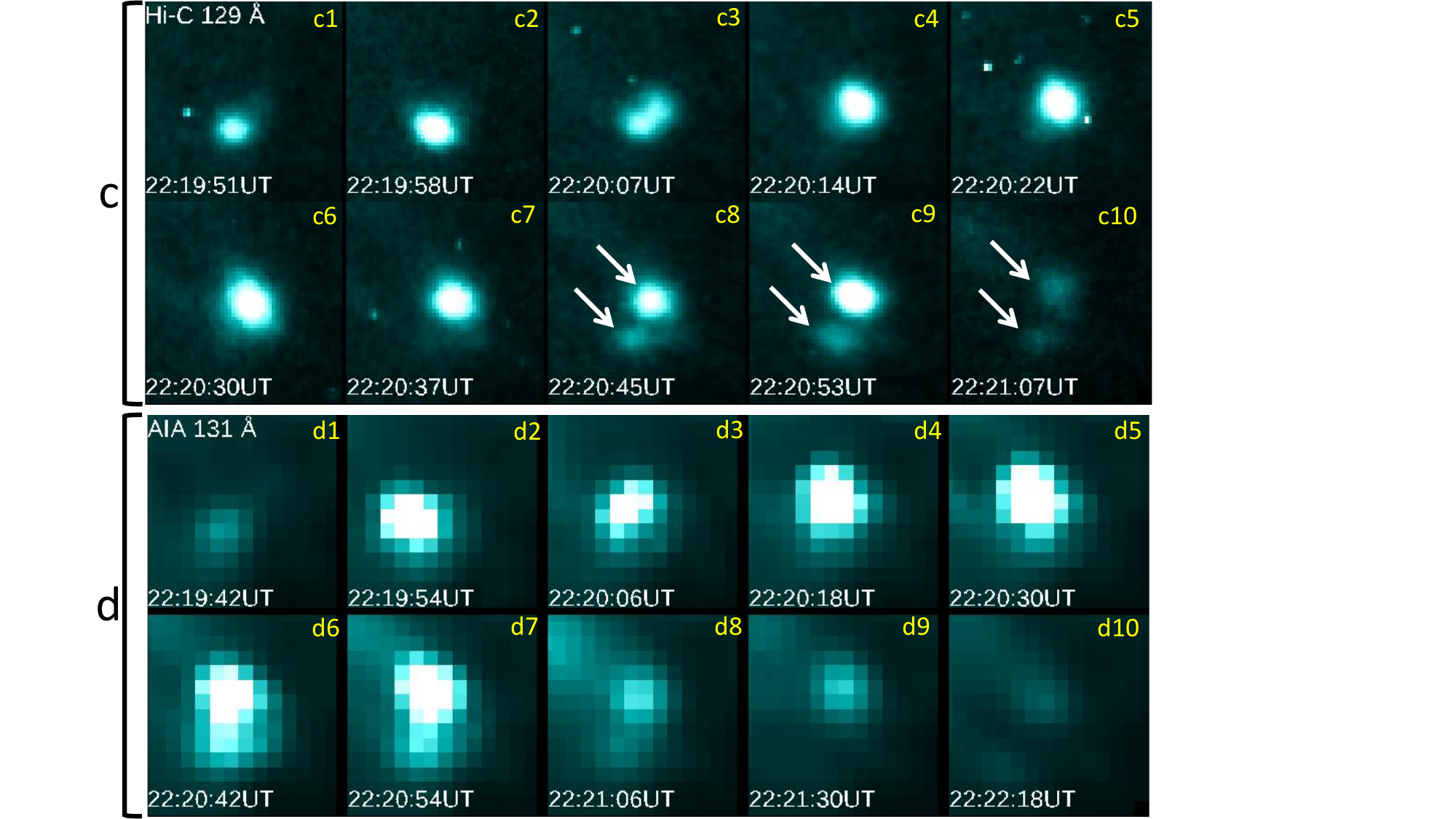}
	\caption{Overview and evolution of the fine-scale bright kernel. Panel (a) contains an image displaying the region of interest of the Hi-C Flare observations. A zoomed in view of the localized brightening is shown in panel (b). An inset in panel (b) presents two plots showing the intensity profiles along two orthogonal cuts through the kernel, as indicated by the white lines, with Gaussian fits over-plotted on the intensity profiles as asterisks. This plot is made to measure the size/diameter of the kernel. The lower two sets of rows, marked by (c) and (d), each contain two rows of Hi-C Flare 129 \AA\ images, and AIA 131 \AA\ images, respectively, showing the evolution of the bright kernel. {The size of each image in panels c and d is 7$^{\prime\prime} \times$7$^{\prime\prime}$.} The difference in appearance of the brightening in panels (c) and (d) arises from the disparity in the spatial resolutions of the two instruments. Note that the timestamps of the respective images on Hi-C and AIA are not identical -- the AIA 131 \AA\ images span a somewhat longer time period. Arrows in panels c8, c9 and c10 mark the splitting  of the kernel into two parts, due to the two loop feet being involved, most clearly visible at 22:20:49 UT in the Hi-C 3 movie. An animation of panel a is available in the online journal, and it runs for seven seconds. 
	}
	\label{fig2}
\end{figure*}

The full field of view (FOV) of Hi-C Flare was 2.2'$\times$4.4', as shown in Figure \ref{fig1}. Because the flare  took place within the region outlined by the red dashed box in Figure \ref{fig1}c, we studied that region in this work.  Figure \ref{fig2}a spans this region of interest. The small red box in Figure \ref{fig2}a centrers on the bright kernel that is under investigation.  A movie of the Hi-C images is available in the online journal (see Figure \ref{fig2}). 

We use most SDO/AIA channels to follow the flare's evolution by creating movies and light curves. Through its seven EUV channels and two UV channels SDO/AIA continuously takes 0.6\arcsec-pixel full-disk images of the solar atmosphere across a broad range of temperature. The cadence is 12 s for EUV channels and 24 s for UV channels.  In this study, we use all the EUV channels: AIA 304, 171, 193, 211, 335, 94, and 131 \AA, which primarily detect emission from plasma at approximately 50,000 K (\HeII), 700,000 K (\FeIX/X), 1.5 MK (\FeXII), 2 MK (\FeXIV), 2.5 MK (\FeXVI), 6--7 MK (\FeXVIII), and 10--12 MK (\FeXXI), respectively. It should be noted that the AIA 94 and 131 \AA\ channels are also sensitive to cooler plasma components, around 1 MK and 0.5 MK, respectively, while the AIA 193 \AA\ channel includes a hotter contribution near 20 MK. In flares, the 171 channel contains a $\approx 20$\% contribution from free–free continuum emission \citep{odwy10}.
Additionally, the AIA 193 and 211 Å channels can detect some cooler plasma as well; see \citet{leme12} for further details. We also use AIA 1600 \AA\ UV images to follow the response of  the lower atmosphere to reconnection-generated high-energy particles from the corona. 
The 1600 \AA\ passband on AIA  primarily captures lower-chromospheric continuum emission but also includes the two C IV lines near 1550 \AA, which form at temperatures around 10$^5$ K in the lower transition region  (TR). 
Short-term brightenings observed in the 1600 \AA\ band have been attributed to these C IV lines,  indicating that such emission originates from the lower TR \citep{leme12}.

After carefully inspecting the AIA images, we found that the despiking process applied during the preparation of AIA Level 1.5 files to remove signals of cosmic particles also removed some small-scale brightenings, including  a portion of the bright kernel that is central to our analysis. To address this, we respiked six AIA EUV wavelengths (131, 94, 171, 193, 211, and 335 Å) and performed our analysis on the respiked images.
{To enable readers to compare the appearance of the bright kernel in despiked and respiked AIA 131 Å images, as well as in other channels, we provide an animation (see Figure \ref{aia_hmi}) that combines 30 minutes of respiked data followed by two hours of despiked data.}
 
\begin{figure*}\centering
	\includegraphics[trim=.1cm 5.8cm .1cm 1cm,clip,width=\textwidth]{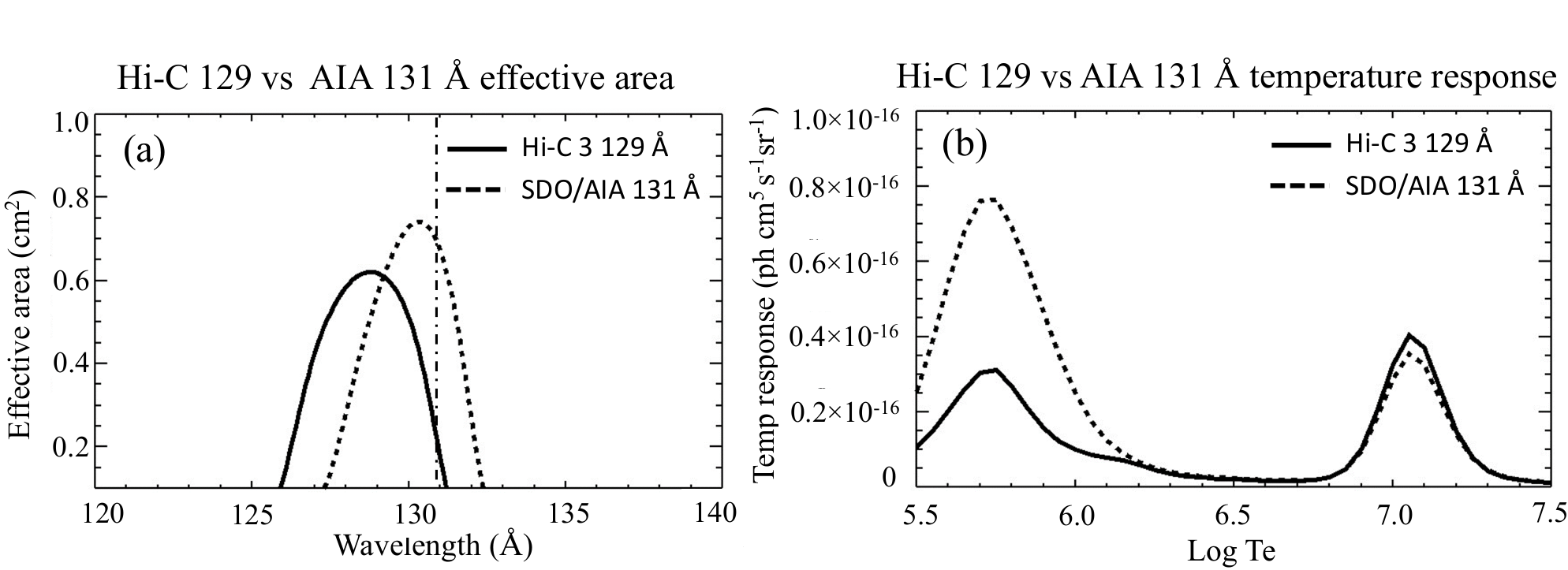}
\caption{Comparison of Hi-C 129 and AIA 131 channel characteristics. Panel (a): The effective area of the Hi-C 3 129 \AA\ passband (solid) and AIA 131 \AA\ channel (dashed). A vertical dash-dotted line marks the wavelength location of Fe VIII ($\sim$131 \AA). Panel (b): The temperature response of Hi-C 129 \AA\ (solid) and AIA 131 \AA\ (dashed). At the cooler temperature peak (log Te = 5.7) AIA is about 2.5 times more sensitive than Hi-C, whereas both are nearly equally sensitive to hotter plasma at the peak of log Te = 7.05.
	}
	\label{fig3}
\end{figure*}

Figure \ref{fig3} compares the effective areas as a function of wavelength for the AIA 131 and Hi-C 129 channels, along with their corresponding temperature responses. It is evident that the AIA 131 channel is more strongly contaminated by cooler emission from Fe VIII compared to Hi-C 129. The temperature response of the Hi-C 129 channel is about 2.5 times less sensitive to cooler emission than that of AIA 131. Therefore, the brightenings observed in Hi-C are more likely from hotter plasma.

The Hi-C Flare campaign included the following instruments: 
Hi-C 3: High-resolution Coronal imager (passband centered at 129 \AA, sensitive to the hot coronal plasma at $\sim$ 11 MK);
COOL-AID: EUV slitless spectrograph centered on 129 \AA;
CAPRI-SUN (high CAdence low-energy Passband x-Ray detector with Integrated full-SUN field of view):  Student-built full-sun soft X-ray detector with rapid cadence  (1000 times faster than GOES); and 
SSAXI-Rocket \citep{moorc24}:  Compact soft X-ray imager.

This {\it Letter} presents a first result from Hi-C 3 data. Data from other Hi-C Flare instruments are still under evaluation and are not considered here.

IRIS and Hinode were observing a different target, as originally planned the previous day/s before the flight, but during the flight the flare occurred in a different AR to which Hi-C was re-pointed. So unfortunately coordinated data from IRIS and Hinode were not obtained for this flare. 

To process the data, first we co-aligned Hi-C Flare 129 \AA\ images and AIA 131 \AA\ images as the passbands for the two are similar (Figure \ref{fig3}). For this we rotated SDO images by about 45 degrees counter-clock-wise and compared the images from Hi-C Flare and AIA 131 and applied any further shifts needed by performing a cross-correlation. Then the other AIA images and the HMI LOS magnetograms were given the same rotation and shifts. 

\section{Results and Discussion} \label{resu}

\begin{figure*}\centering
\includegraphics[trim=0cm 0cm 0cm 0cm,clip,width=\textwidth]{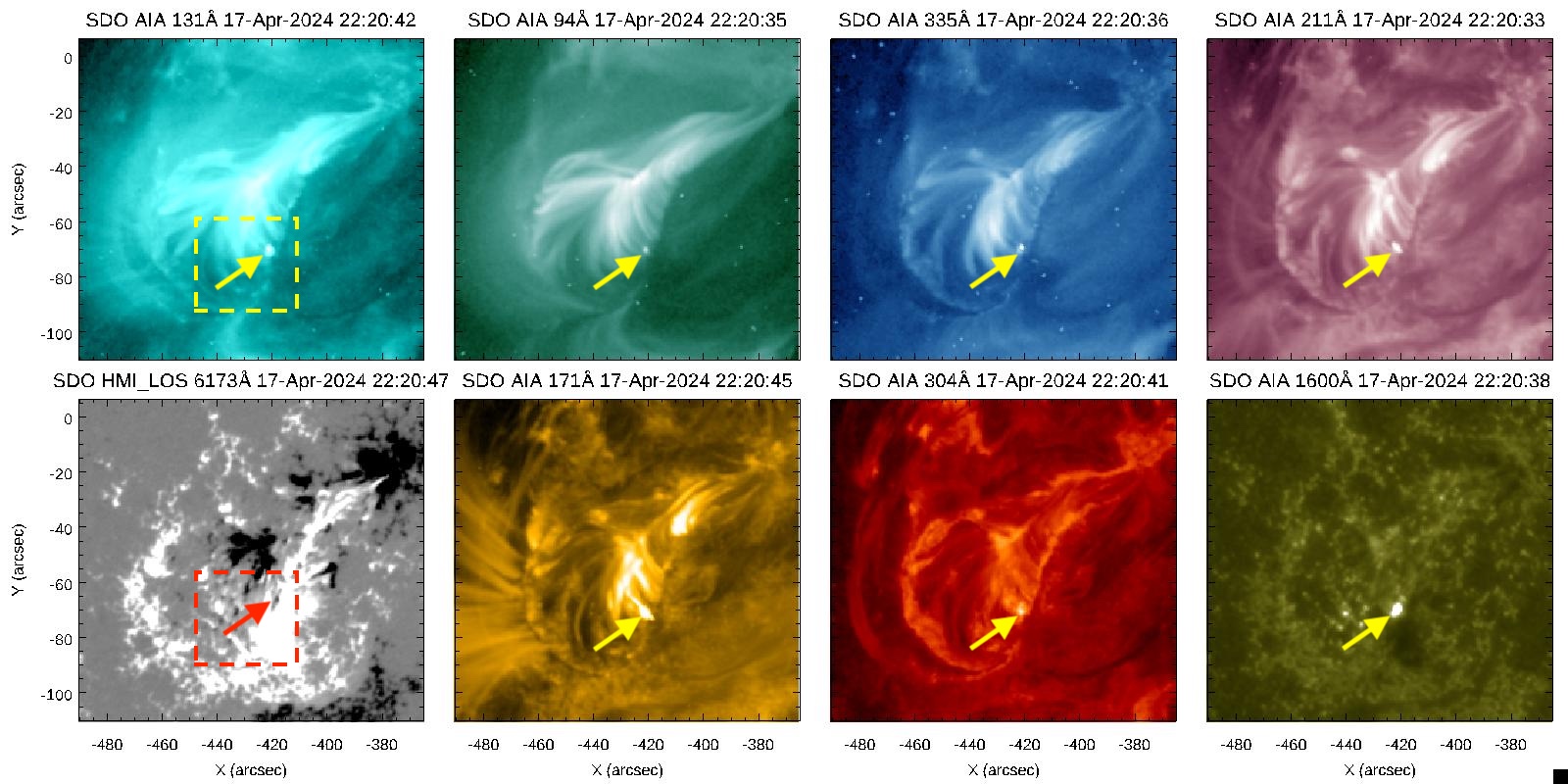}
	\caption{SDO observations of the M1.6-class flare are shown in various AIA channels, along with the HMI line-of-sight magnetogram. An arrow in each panel points to the location of the bright kernel under investigation. A boxed region covers two locations of similar brightenings: one captured by both Hi-C Flare and SDO, and the other observed only by SDO, approximately six minutes after the Hi-C Flare observations concluded. Both brightenings are shown in Figure \ref{fig4}. {An animation displaying all these panels, first showing the flare in respiked AIA images for about 30 minutes and then in despiked images for about two hours, is available in the online journal. The total duration of the animation is 1 minute and 15 seconds.}
	}
	\label{aia_hmi}
\end{figure*}

\begin{figure*}\centering
\includegraphics[trim=0cm 0cm 0.5cm 0cm,clip,width=0.91\textwidth]{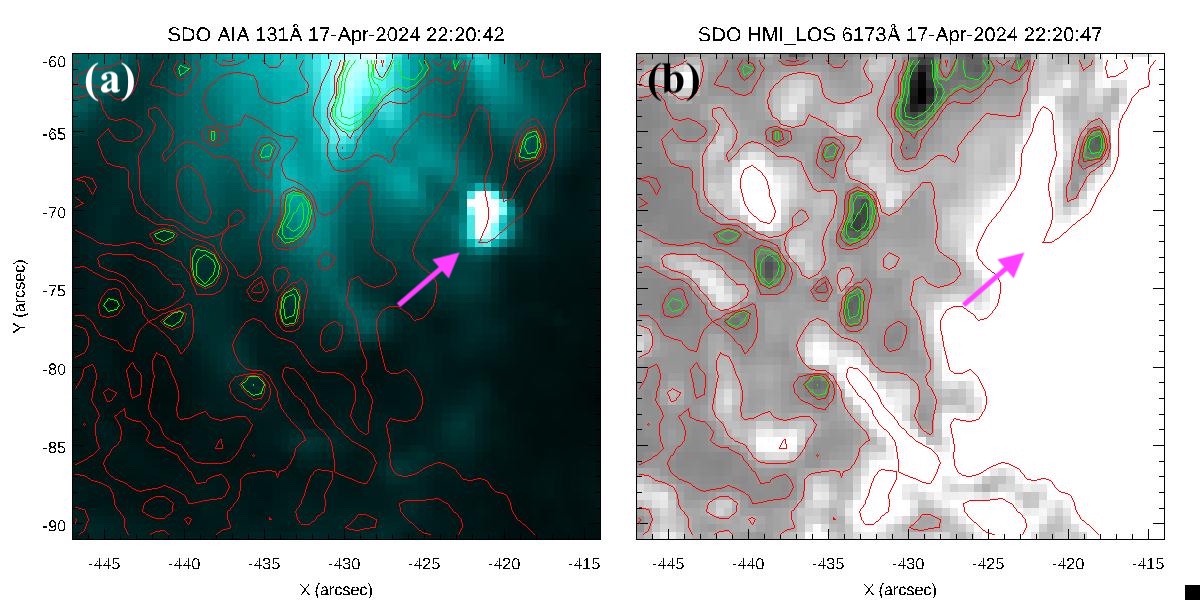}
	\includegraphics[trim=0cm 0cm 0.5cm 0cm,clip,width=0.91\textwidth]{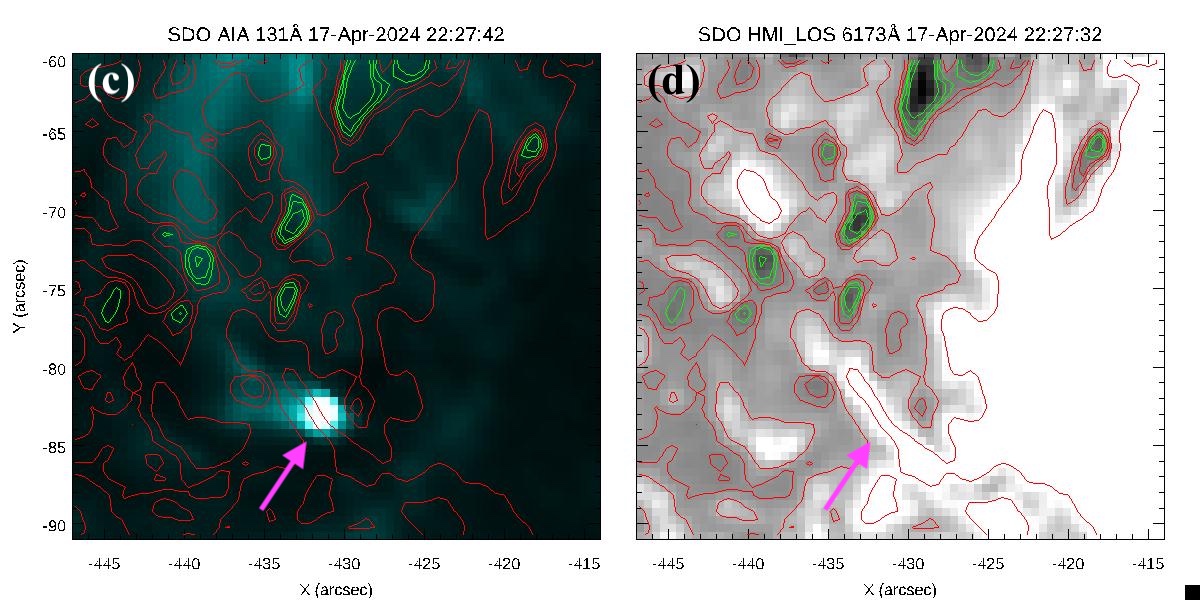}
	\caption{Magnetic field distribution at and near two bright kernels. Panel (a): SDO/AIA 131 \AA\ image during the localized brightening activity - this event is captured by Hi-C Flare. An arrow in pink points to the bright kernel. The corresponding HMI LOS magnetic field is shown on the right panel (b), displaying the magnetic field distribution. The red and green contours, also overplotted on panel (a) correspond to positve and negative polarity, respectively, with contour levels of [$\pm20, \pm 60, \pm150, \pm500$] G.  The bottom panels (c) and (d) are the same as the top two but for a different time (about 5-minutes later than the top ones) showing a second similar brightening event, outside the Hi-C observation time, and so only captured by SDO. The dashed yellow or red box in each of Figure \ref{aia_hmi} panels AIA 131, and HMI LOS magnetogram, outlines the FOV of these images. An animation is available in the online journal, with a total duration of 15 seconds.}
	\label{fig4}
\end{figure*} 

\begin{figure*}\centering
	\includegraphics[trim=0cm 0cm 0cm 0cm,clip,width=0.82\textwidth]{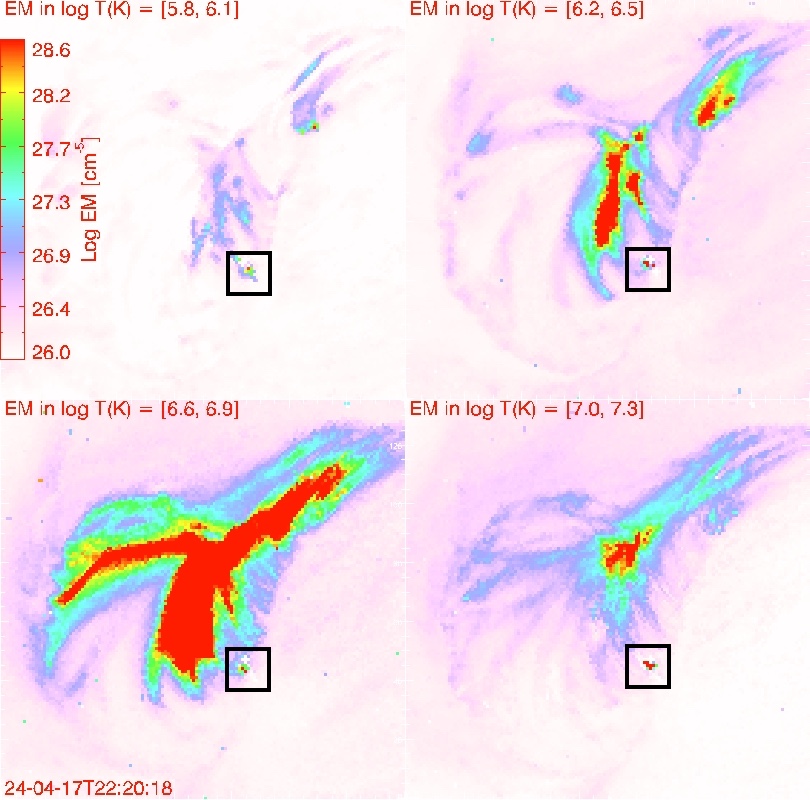}
\caption{Emission measure (EM) maps created from six AIA channels [94, 131, 171, 193, 211, 335], displayed in four log temperature (log$_{10}$T) bins for a given time (22:20:18). The black box in each panel centers on the bright kernel. The kernel carries significant EM in all temperature bins, thus representing its multi-thermal nature. An animation is available in the online journal, with a total duration of 15 seconds. 
	}
	\label{dem_map}
\end{figure*}

Figure \ref{fig1} provides an overview of the Hi-C Flare observations, showing the full field of view (FOV) recorded by Hi-C and the FOV of Figure \ref{fig2}a, and Figure \ref{aia_hmi} . The accompanying GOES X-ray plots summarize the M1.6-class flare and indicate the time interval of the flare segment captured by Hi-C. The distribution of the flare loops suggests the flare occurred in a fan-spine topology \citep[e.g.,][]{masson09, sun13, janv16}. The flare loops follow a distinct dome shape, rooted in a semi-circular ribbon, and extend along a spine rooted at $X \approx -370\arcsec$, $Y \approx -20\arcsec$.

{Figure \ref{fig2} shows the Hi-C region covering the flare, along with the localized brightening, the prospective bright flare-ribbon kernel, analysed in detail. Regardless of its origin (whether a microflare or a localized flare-ribbon kernel, unless otherwise specified), we refer to this brightening simply as the bright kernel.} Near its peak intensity, the kernel’s diameter is estimated to be 875 $\pm$ 25 km, obtained by averaging its horizontal and vertical extents as illustrated in Figure \ref{fig2}b. The {measurement} uncertainties in each direction were determined by measuring the diameter along three adjacent lines of pixels along the two orthogonal cuts shown in Figure \ref{fig2}b. Since the two orthogonal extents are similar and the kernel appears roughly circular, we averaged the two measurements and their associated uncertainties to define its diameter. {The measured diameter represents an upper limit, and the true size of the kernel could be smaller. In particular, any substructure on scales below the telescope's PSF ($\simeq$0.45"; \cite{koba14,rach19,vigi25}) would not be resolved in these observations.}
The size of the bright kernel is comparable to that of typical flare ribbon kernels. For example, \citet{roma17} reported that circular ribbons in flares are composed of several adjacent compact kernels with sizes of about 1" -- 2". We note that smaller flare ribbon kernels are detected in higher resolution observations, \citep[e.g.,][]{jona25,sing25,yada25}. 
  
The evolution of the kernel in the Hi-C Flare 129 \AA\ images (Figure \ref{fig2}c) and their light curves (described later) reveal that the kernel lived for 90 $\pm$ 1.3 seconds.
In Hi-C and SDO movies, there is evidence of magnetic field restructuring in the corona associated with the kernel {(compare Hi-C image frames for example at 22:18:40 and 22:19:51 in the Hi-C movie accompanying Figure \ref{fig2}), }suggesting coronal magnetic reconnection during the brightening. Furthermore, the Hi-C brightening showed a subtle north-westward motion between 22:20:00 and 22:20:15~UT, with a plane-of-sky speed of about 50~km~s$^{-1}$. These observed dynamics align with the apparent motions of flare kernels attributed to slipping reconnection \citep[e.g.,][]{dudi14, li15, lori19}.

During its evolution, the kernel splits in two, particularly at the end (see the panels c8--c10). In Figure \ref{fig2}d, the evolution of the kernel is also shown in the AIA 131 \AA\ channel, the AIA channel closest to the 129 \AA\ of the Hi-C Flare images.  The AIA 131 images also show splitting of the kernel in two at the corresponding times (see the panels d7 and d8).  When viewed closely in the SDO movie (Figure \ref{aia_hmi}), there appears to be two upward-reaching coronal loop legs (see in the movie frames near time 22:20:30 UT in AIA 131, 94, 211 and 171 \AA), one rooted in each of these two kernels. Apparently when the whole kernel was very bright the two parts appeared to be merged and looked like a single kernel, and with time during cooling the two loop feet separate into the two bright kernels. Perhaps these two kernels and their corresponding loop legs are a result of the evolution of the magnetic field. These two loop feet can be noticed separated in the Hi-C 129 \AA\ movie (see near 22:20:30 UT, and at 22:20:49 UT), as well as in the AIA 131 \AA\ image shown in Figure \ref{fig4}a. 

Splitting and merging kernels have been seen before in flare ribbons in the beginning of a C-class solar flare in Ca II H interference filter images centred on 3968 \AA\ by \cite{sobo16} using high-resolution  GREGOR data. As a result of the splitting and merging interrupting the tracking, these ``knots" were not followed during their whole lifetime. Their tracking periods ranged from 32 s to 86 s for different knots having lifetimes of about a minute. These values are similar to the lifetime of our bright localized kernel, and their knots show splitting and merging, like that in our case. Similar footpoint motions were observed by \cite{qiu02} in UV and EUV images, but again in flare kernels in the beginning of a solar flare. 

We note that \citet{czay99} were the first to report chromospheric evaporation in transition-region and coronal lines during the late gradual phase of a flare. Their conclusion, based on the observed upflows and downflows, implied that reconnection was ongoing throughout their flare observations. Flare-ribbon kernels extending over a sunspot umbra were later reported by \citet{sobo16}. In contrast to our isolated bright kernel, both of these studies involved fully developed flare ribbons during the decay phase of M- and C-class flares, respectively.

A second localized brightening occurs slightly south-east of the above kernel about five minutes later.  This second kernel was not captured by Hi-C Flare but it shows striking similarity with the first one in SDO data. 
In Figure \ref{fig4} we display SDO/AIA 131 \AA\ images with LOS magnetic field contours on them, together with the corresponding contoured LOS magnetograms. The FOV covers both kernels. 
This Figure shows that each kernel is located near the edge of a negative polarity patch in a dominant positive magnetic polarity region.  The neutral line nearest the kernel in each case is not sharp, and Hi-C and AIA images show no obvious extensions toward the opposite polarity magnetic flux patch. This suggests the origin of this fine-scale heating event may not be by magnetic reconnection in the lower atmosphere that often accompanies flux cancellation, as seen for small-scale dots, tiny loops, and small surges by \cite{tiw19}, and small-scale jets by \cite{pane19} using data from Hi-C 2.1 \citep{rach19}. Nonetheless, given the nearness to the neutral line that cannot be strongly concluded. For an example,  the transition region images from IRIS in \cite{tiw19} show an extended intensity structure reaching to the smaller bright dot-like heating events seen in Hi-C2.1 172 \AA\ images.

Somewhat similar coronal brightenings above flux cancellation were also recently reported in the synthetic images of Si IV 1400 \AA\ of a Bifrost MHD simulation \citep{tiw22}. In Fe IX/X images, the brightenings were smaller than the corresponding extended structures in Si IV synthetic images. It was concluded that only the hottest counter-part of magnetic reconnection in the lower atmosphere showed up in Fe IX/X emission, while the lower more dense atmosphere showed larger structures in Si IV line.
Unfortunately, in the absence of coordinated IRIS observations for this flare, we cannot determine whether a similar extended structure was present in the transition region, and therefore cannot test the possibility of magnetic reconnection as the cause of the localized fine-scale hot kernel.

\begin{figure*}\centering
	\includegraphics[trim=0.71cm 11.6cm 1cm 4cm,clip,width=0.64\textwidth]{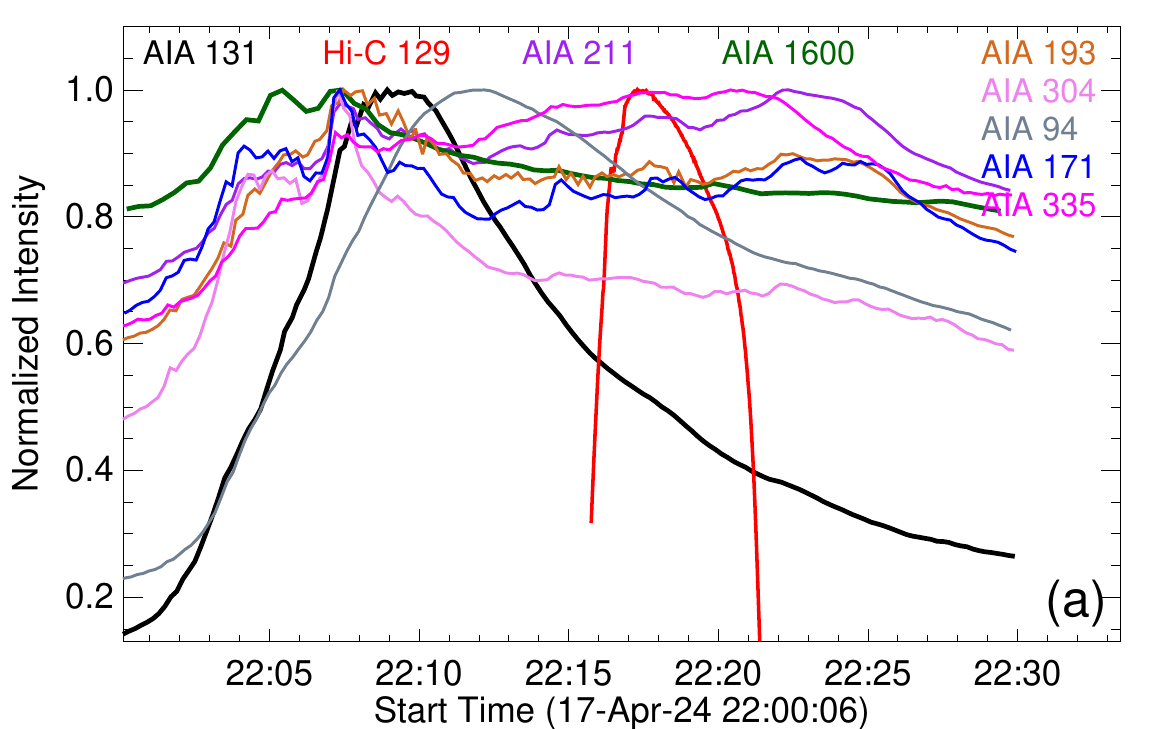}
	\includegraphics[trim=2cm 0cm 1.55cm 0.9cm,clip,width=0.26\textwidth]{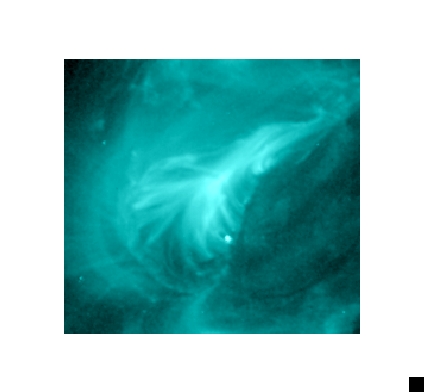}
	\includegraphics[trim=0.71cm 11.6cm 0.89cm 4cm,clip,width=0.64\textwidth]{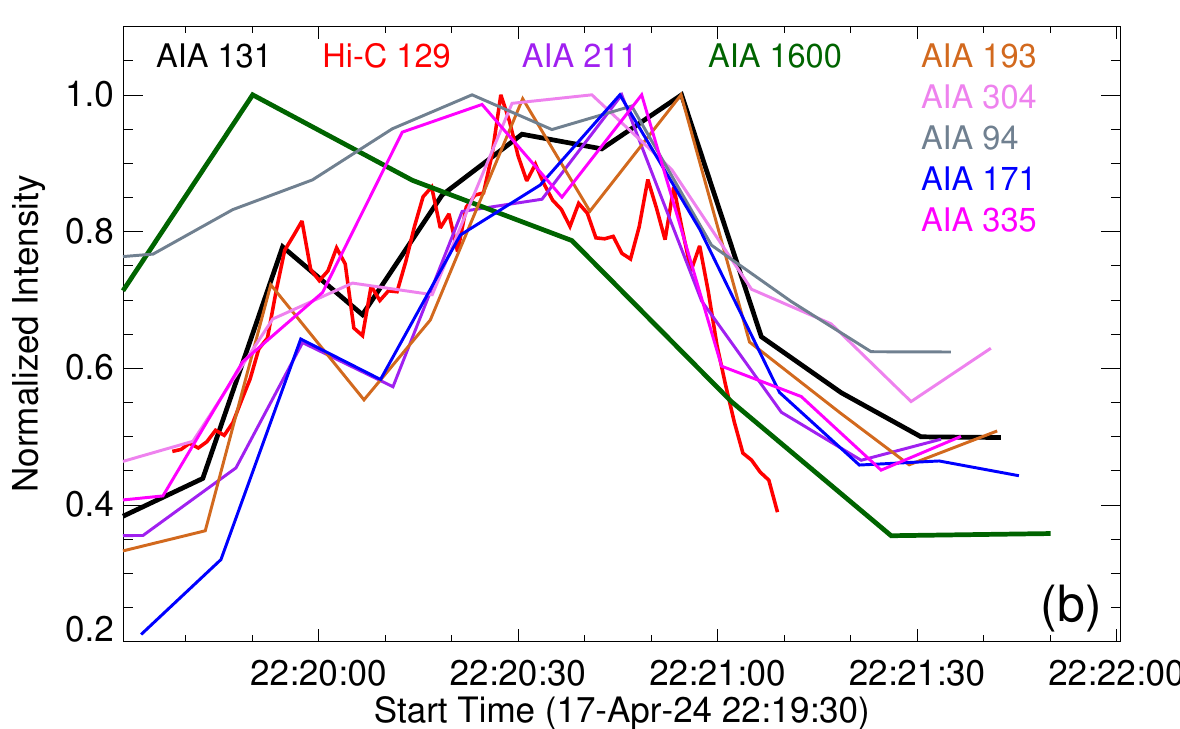}
	\includegraphics[trim=2cm 0cm 1.55cm 0.9cm,clip,width=0.26\textwidth]{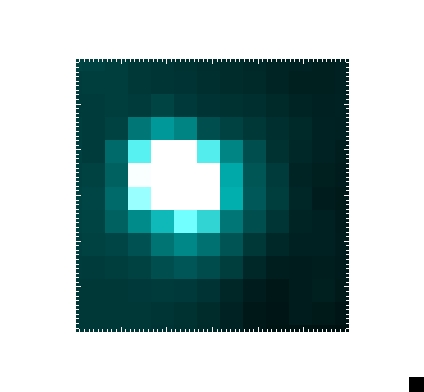}
	\includegraphics[trim=0.71cm 11.6cm 1cm 4cm,clip,width=0.64\textwidth]{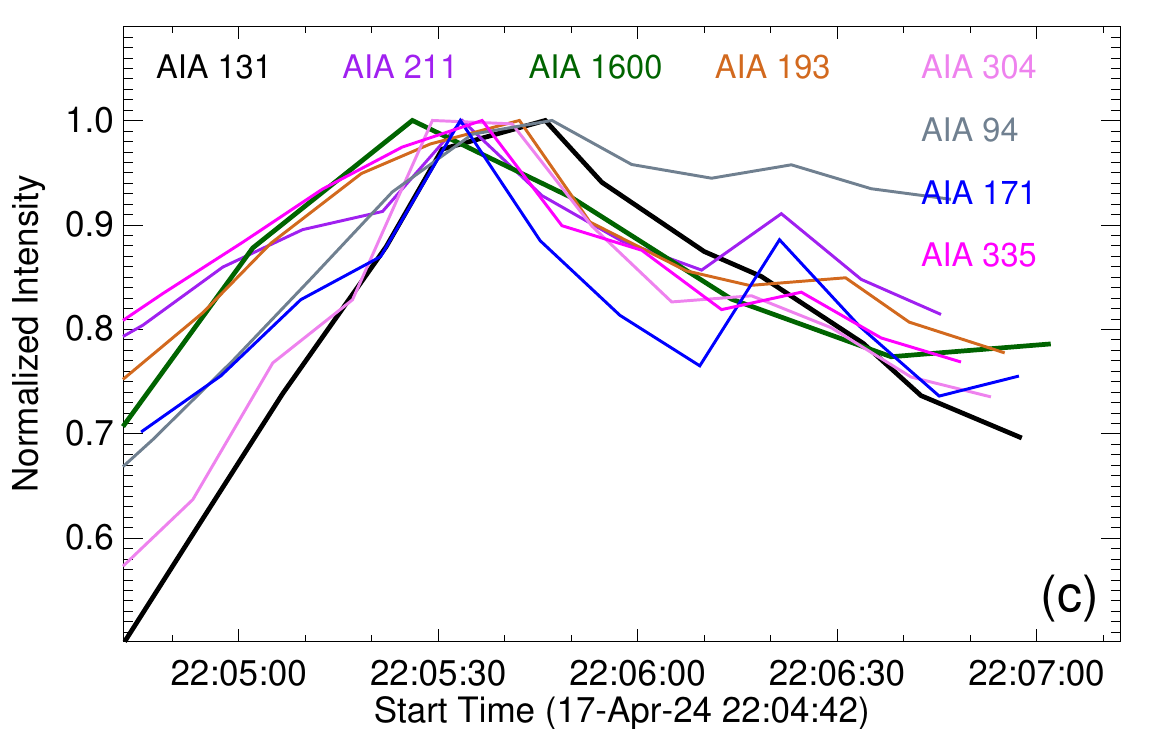}	
		\includegraphics[trim=2cm 0cm 1.55cm 0.9cm,clip,width=0.26\textwidth]{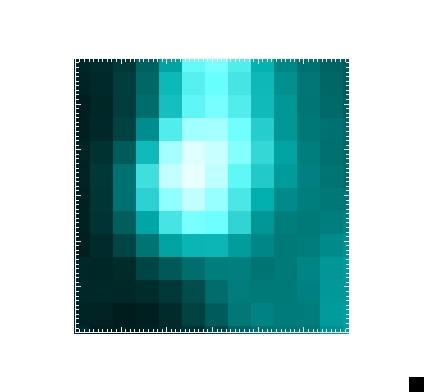}
	\caption{Light curves from SDO/AIA and Hi-C Flare data. Panel (a) shows light curves for the full FOV shown in Figure \ref{fig2}a. Panel (b) shows light curves for the bright kernel in the FOV shown in Fig \ref{fig2}b. Panel (c) shows light curves of different AIA channels of a localized flare-ribbon kernel during the impulsive phase of the flare. Insets in the right to each panel shows the region, each in AIA 131 channel, for which the corresponding light curves were created.
		Each light curve in the figure is colored to match its corresponding wavelength annotation or legend. The light curves of 1600 and 131 are made thicker for easy distinction. 
}
	\label{lightcurves}
\end{figure*}

The emission measure (EM) maps in Figure \ref{dem_map}, calculated using six AIA channels [94, 131, 171, 193, 211, 335] using the method described in \cite{cheu15dem}, reveal that the bright kernel is multi-thermal in nature. 
	{This is consistent with previous studies reporting multi-thermal emission at flare loop footpoints, for example manifested by multi-peaked DEM distributions \citep[e.g.,][]{delz11flare,grah13,kenn13}.}

Somewhat similar hot, localized bright kernels were reported by \cite{mill09} using Hinode/EIS Fe XXIII and Fe XXIV lines (formed at $>$ 12 MK) at a loop footpoint, but again during the impulsive phase of a C-class flare. To the best of our knowledge, such hot, localized bright kernels during the post-maximum phase of a solar flare have not previously been reported.

To find out if the bright kernel gives luminosity curves similar to flare-ribbon kernels, we created light curves for different AIA channels together with the Hi-C 129 \AA\ light curves. Figure \ref{lightcurves}(a) shows light curves for the larger FOV, covering the whole active region, for reference. As is usual in solar flares, AIA 1600 \AA\ peaks first displaying the first response to impulsive down-streaming charged particles from reconnection. 
Furthermore, AIA 1600 Å shows two peaks, likely corresponding to two successive streams of charged particles.
The AIA 304, 171, 193, 211, and 335 Å channels also exhibit intensity enhancements nearly concurrent with the two 1600 \AA\ peaks. Later, the 131 Å channel peaks, then the 94 Å channel, and then consecutively later  luminosity enhancements in the 335, 211, 193, and 171 Å channels, that is, the normal cooling sequence for flares, is seen. The first and second peaks in AIA 1600 \AA\ luminosity occur approximately 300 s and 150 s before the peak of the 131 Å luminosity, respectively.{ We note that the uncertainty in SDO/AIA pixel EUV brightness, which depends on the gain of the CCD, is significantly smaller than the individual peaks of the light curves. Please see Table 6 of \cite{boer12} for the camera-CCD gain and read noise.}

The panel (b) of Figure \ref{lightcurves} shows light curves of the small box covering the bright kernel in Figure \ref{fig2}. Similar to the light curves during the impulsive phase of the flare, the localized brightening curves show that 1600 Å peaks first. Immediately afterwards (within about 30–50 s), Hi-C 129 Å and all other AIA channels reach their peak intensities. Because the FOV does not include the loops, a subsequent cooling sequence is not observed. 
We judge that these light curves are consistent with  this localized brightening being the footpoint of an isolated new flare loop. The AIA 1600~\AA\ luminosity peaks approximately 50--60~s earlier than the AIA 131~\AA\ luminosity peak. Such early peaking of the 1600 Å emission in solar flares, seen in panels (a), (b) and (c) of Figure \ref{lightcurves}, has been well observed by \citet[][and references therein]{qiu25}.

 Panel (c) in Figure \ref{lightcurves} shows AIA light curves of a flare-ribbon kernel during the impulsive phase of the flare. Similar to the light curves in the bright kernel observed by the Hi-C, the 1600 channel peaks first. However, the time difference between the 1600 peak and the subsequent 131 peak is shorter (by about 25 s) than that observed for the bright kernel in panel (b).{ This is the case for several of the bright kernels inspected -- 1600 peaks about 25 s before 131 does.} Nonetheless, this observation supports the interpretation that the bright kernel under study is indeed an isolated, localised, flare-ribbon kernel. A caveat is that a local reconnection burst can also produce such a delay, as the current sheet may extend to higher altitudes and heat less dense plasma, resulting in a similar scenario.

There is the presence of similar localized brightenings at multiple occasions before the flare started, also previously reported in other solar flares, e.g., by \citet{josh11Bhuwan}; see, for example, the movie accompanying Figure \ref{aia_hmi}, at and around AIA 131 timings of 21:22:54, 21:28:54, 21:35:18, 21:36:42, 21:40:54, 21:42:54, 21:52:42, but only one (at 22:44:54)  is seen after the flare has ended. Because the magnetic field distribution on the photosphere and field's general configuration remains somewhat similar, as seen in the movie accompanying Figure \ref{fig4}, we judge that these localized brightenings are most likely a result of streamed charged particles from a coronal magnetic reconnection. After the flare has ended the field is more potential and so local coronal reconnection events cease. Further, the light curves of similar bright kernels during the impulsive phase of the flare show a similar behaviour as the light curves for the localized bright kernel seen by Hi-C. 

While localized small-scale flare kernels, in flare-ribbons, sometimes referred to as `knots', as mentioned previously, are common and expected per standard flare models to occur during the initial phase of a solar flare \citep{shar14,sobo16,youn24,jona25,yada25}, their presence during the decay phase in EUV observations has not been reported before. 
\cite{shar14} explained their observed fine structuring of flare ribbons by Joule dissipation if the electrical resistivity is enhanced in the partially ionized plasma of the lower solar atmosphere, or by turbulent resistivity due to even smaller structuring. These mechanisms perhaps apply to the localized bright kernel we observe here in the  post-maximum phase of the flare. 

We note that, although not previously reported, isolated bright kernels during a flare's gradual phase may not be uncommon; however, contamination from cooler emission can reduce their contrast, making them harder to detect in AIA 131 Å. This feature might have been less apparent without the Hi-C data (see Figure \ref{fig3}).

Our inference that the brightening is a flare kernel in late phase is consistent with the idea of prolonged reconnection for a prolonged heating in solar flares  \citep{czay99,czay01,warr06_flare,reev07,qiu12,reep20}. Bursty reconnection \citep[e.g.,][]{naka09} in coronal loops after the main phase of a flare plausibly makes localized flare-ribbon bright kernels in the late phase. We think this is what is happening in our current observation.  

In a time–distance map (not shown), we found no evidence of plasma downflows along the loop leg  to the bright kernel. This indicates that the localized brightening was unlikely caused by impact compression from a downflow \citep{reev17}.

The present Hi-C Flare observations raise several questions: How common is the occurrence of isolated flare kernels during the post-maximum phase? Is it more prevalent in M-class flares than in other classes, such as X-class or C-class flares? Why is the brightening so highly localized? Why did neighbouring coronal loops not participate in the reconnection, forming an extended flare-ribbon as seen in the impulsive phase?
Future MHD simulations should seek to account for the presence of such fine-scale kernels during the flare  post-maximum phase. 
Perhaps a flare's impulsive/rise and maximum phases are a storm of fine-scale bursts of reconnection in the flaring coronal magnetic field \citep{lin84}, each burst being similar to the burst that made either of the two bright kernels reported here. 

\subsection{A ``late-phase localized flare ribbon kernel" or a ``microflare"?}

Based on the discussions in Section \ref{resu}, here we make a list of observational support to each of the two scenarios.

\subsubsection{Support for the kernel being a localized flare-ribbon kernel caused by coronal magnetic reconnection} \label{support_kernel}

\begin{itemize}
	\item	As noted earlier in Section \ref{resu}, many similar localized brightenings occurred at or near the same location as the brightening under study, prior to the flare. See, for example in the movie accompanying Figure \ref{aia_hmi}, the AIA 131 \AA\ images at and around 21:22:54, 21:28:54, 21:35:18, 21:36:42, 21:40:54, 21:42:54, 21:52:42. But only one such brightening (at 22:44:54) is observed after the flare has ended. Since the photospheric magnetic field distribution remains largely unchanged (see the movie accompanying Figure \ref{fig4}), we infer that these localized brightenings are likely caused by streams of charged particles from coronal magnetic reconnection. After the flare, the field becomes less non-potential, and coronal reconnection ceases.
	
	\item The localized brightenings do not show a preference for mixed-polarity regions, as would be preferred by a micro-flare event. Several of the brightenings noted above are seated at unipolar magnetic regions, as is characteristic for flare ribbons.  

\item Further, the light curves of similar bright kernels during the impulsive phase of the flare show a similar behavior to the light curves for the localized bright kernel seen by Hi-C. The uncertainty in pixel EUV brightness, which depends on the CCD gain, is much smaller than the peak intensities in the light curves. This readout error is particularly low for AIA \citep{boer12}. We verified that, for all AIA channels, the uncertainties are significantly smaller than the observed peaks in the light curves.

\item There is no \textit{sharp} polarity inversion near the bright kernel.  We think that favors that the kernel was made by some mechanism other than local reconnection in a microflare.  

\item That the kernel is at the foot of a loop leg that extends toward the spine of the active region suggests its brightening results from reconnection in the corona.  Evidence of magnetic field restructuring in the cusp of flare loops that include the loop rooted in the kernel also favors this interpretation.

\item The bright kernel is also unlikely to be a small bipole without coronal temperatures, as for example as seen in 1600 \AA\ by \cite{tiw14}. As previously mentioned, the magnetograms do not show an obvious polarity inversion line, or both polarities of a bipolar system. Further, Figure \ref{fig3} shows that Hi-C 129 is less sensitive to cooler plasma than AIA 131; therefore, its clear visibility in Hi-C supports the interpretation that the bright kernel has coronal temperatures.

\end{itemize}

\subsubsection{Support for the kernel being a localized reconnection micro-flare event}\label{support_microflare}

\begin{itemize}
	\item  The bright kernel is much brighter than other similar localized brightenings, and so even if others may be generated by coronal magnetic reconnection, this kernel could be caused by a local magnetic reconnection.
	
	\item In previous observations of Hi-C 2.1 \citep{tiw19} and Solar Orbiter's EUI/HRI as well as in simulations \citep{tiw22}, an extended structure like jet spire from the reconnection site has been observed in transition region temperatures. In the absence of IRIS observations a scenario of local magnetic reconnection, e.g., producing a jet from the kernel, cannot be ruled out. 
	
	\item A sharp polarity inversion line might not be necessary for a local-reconnection microflare to occur. Magnetic reconnection can happen as a standing current sheet, starting from or extending to the higher atmosphere \citep{hans19}.
	
	\item In the HMI magnetograph, scattered light might hide few-pixel clumps of opposite-polarity flux that are inside a larger area of unipolar flux as strong as that surrounding our kernel (see, e.g., \cite{wang16_loop,pane18a,pane19}).
	
	\item  There is no obvious conjugate footpoint to the localized bright kernel. However, a conjugate flare loop footpoint may sometimes be too spread out to be distinctly visible, particularly in such a localized coronal reconnection event as ours. 
	Alternatively, the conjugate footpoint may sometimes get obscured by flare emission in the corona along the line of sight.
	Finally, not all flare-ribbon kernels have a conjugate kernel that shows corresponding intensity and velocity evolution \cite[see, e.g., Section 3 of][]{sobo16}.
	
\end{itemize}

In our view, these points favor interpreting the bright kernel as a flare ribbon kernel produced by a burst of coronal magnetic reconnection. Nonetheless, the arguments in Sections \ref{support_kernel} and \ref{support_microflare} do not definitively confirm either scenario.

\section{Conclusions}\label{conc}

We report the observation of a fine-scale, {localized bright kernel} during the post-maximum phase of an M1.6-class solar flare, captured in 129 \AA\ hot emission ($\sim$ 11 MK) by the first Hi-C Flare rocket flight experiment (the third successful Hi-C flight). This brightening lasted for about 90 $\pm$ 1.3 seconds and had a diameter of {at most} 875 $\pm$ 25 km. SDO/AIA 131 \AA\ images reveal a second, similar brightening approximately five minutes later at another location. { In SDO/HMI LOS magnetograms each of these bright kernels is in positive unipolar flux but near an inclusion of negative flux — suggesting the possibility of a burst of local magnetic reconnection being the cause. However, various observational facts favor the bright kernel being a localized bright flare-ribbon kernel produced by accelerated charged particles from coronal magnetic reconnection. }

Several qualitatively-similar brightenings were observed in the same or nearby locations before and during the flare. Such activity in this region, on the other hand, declined after the flare gradual phase. This suggests that these localized brightenings were manifestations of a response of the lower atmosphere to heating driven by flare reconnection acting in the conversion of free energy accumulated in the corona.
All of these brightenings, including the one observed by Hi-C, were located at the footpoints of flare loops, the defining characteristic of flare kernels. These flare loops constitute the dome of the fan-spine topology, directed towards the likely location of a null point at the dome apex. Furthermore, the Hi-C brightening exhibited an inconspicuous motion towards the north-west between 22:20:00 - 22:20:15\,UT, at the plane-of-sky speed of about 50\,km\,s$^{-1}$. Such dynamics are consistent with the apparent flare kernel motions driven by slipping reconnection \citep[e.g.,][]{dudi14, li15, lori19}. And finally, the AIA 1600 \AA\ light curves peaked about 50 seconds before the 131 \AA\ emission. Inspection of light curves of several flare ribbon kernels reveals a similar trend in that 1600 Å peaks about 25 s before 131 Å peaks.
The arguments above provide further evidence that the structure observed by Hi-C is a late-phase isolated flare-ribbon kernel, resulting from coronal magnetic reconnection and the consequent impact of accelerated charged particles on the lower atmosphere.

{We note however that the transition region emission dominating the 1600 \AA\ channel in flares responds very rapidly to flare heating from high altitudes irrespective of the heating agent \citep[see, e.g.,][]{lori25}, i.e, whether the heating is by non-thermal particles, heat conduction,  or by waves from the burst of coronal reconnection, or by a local burst of reconnection (microflare) low in the local magnetic field.}

This {\it Letter} gives a first report of a {prospective} isolated, localized, multi-thermal, flare-ribbon kernel in the late phase of a solar flare. As each flare is unique, the prevalence of this phenomenon remains to be determined.
Future Hi-C Flare flights and similar observations, ideally coordinated with IRIS, Hinode, Solar Orbiter, MUSE, and DKIST, should provide decisive insight into the cause of these localized flare kernels across different solar flares, thereby establishing their role in the post-maximum phase, and the entire flare process. 
  
\acknowledgments
{We thank the referee for their insightful comments, which have helped improve the presentation of this paper. We also express our gratitude to Paola Testa and Meng Jin for their valuable discussions on the readout error estimates for the AIA channels.} S.K.T., N.K.P., and R.L.M. gratefully acknowledge support by NASA HSR grant (80NSSC23K0093), and NSF AAG award (no. 2307505). S.K.T. also aknowledges support by NASA contract NNM07AA01C (Hinode). NKP acknowledges support from NASA's SDO/AIA grant (NNG04EA00C) and HSR grant (80NSSC24K0258). R.L.M acknowledges the support from the NASA HGI program. J.L., and V.P. acknowledge support from NASA's HGI grant 80NSSC24K0553. J.L., V.P., and B.D.P. gratefully acknowledges support from NASA contract NNG09FA40C (IRIS). B.D.P. was supported by grant 80NSSC23K0164. We acknowledge the High-resolution Coronal Imager (Hi-C Flare) instrument team for making the flight data available under NASA:Heliophysics Flight Opportunities for Research and Technology (HFORT) Low Cost Access to Space (LCAS) program (proposal HFORT19-0025).  MSFC/NASA led the mission with partners including the Smithsonian Astrophysical Observatory, the University of Central Lancashire, and Lockheed Martin Solar and Astrophysics Laboratory.  Hi-C Flare was launched on 2024 April 17. 
The AIA and HMI data are courtesy of NASA/SDO and the AIA and HMI science teams.

\end{document}